\documentclass[aps,article,twocolumn]{revtex4}
\usepackage{graphicx}
\usepackage{amsmath}
\usepackage{bm}

\begin{document}

\title{Engineering of radiation of optically active molecules with chiral nano-meta-particles}

\author{Vasily V. Klimov$^{1*}$, Dmitry V. Guzatov$^2$, Martial Ducloy$^3$}

\affiliation{$^1$P.N. Lebedev Physical Institute, Russian Academy of
Sciences, 53 Leninsky Prospect, 119991 Moscow, Russia \\ $^2$Yanka
Kupala Grodno State University, 22 Ozheshko str., 230023 Grodno,
Belarus and \\ Laboratoire de Physique des Lasers, UMR CNRS 7538,
Institut Galilee, Universite Paris-Nord, Avenue J-B. Clement, F
93430 Villetaneuse, France}

\date{\today}
\maketitle

\textbf{The radiation of optically active (chiral) molecule placed near
chiral nanosphere is investigated. The optimal conditions for
engineering of radiation of optically active (chiral) molecules with the
help of chiral nanoparticles are derived. It is shown that for this
purpose, the substance of the chiral particle must have both
$\varepsilon$ and $\mu$ negative (double negative material (DNG)) or
negative $\mu$ and positive $\varepsilon$ ($\mu$ negative material
(MNG)). Our results pave the way to an effective engineering of
radiation of ``left'' and ``right'' molecules and to creating pure
optical devices for separation of drugs enantiomers.}

It is well known that nanoparticles influence substantially both
fluorescence of molecules and Raman scattering of light by
molecules. These effects are especially strong in the case of
metallic nanoparticles where plasmon resonances can be excited. As a
result, it is possible to detect radiation even from single molecule
(surface enhanced Raman scattering SERS \cite{ref1} and surface
enhanced fluorescence SEF \cite{ref2}) and to use this effect in
various applications, in particular, for detection of certain
proteins in disease diagnosis \cite{ref3}. Recently it was shown
that planar chiral metamaterials \cite{ref4} and even nonchiral
plasmonic nanoparticles \cite{ref5} are able to dramatically (by
several orders of magnitude) change the chiral dichroism of a chiral
molecule and thus such metamaterials could form the basis for
assaying technologies capable of detecting amyloid diseases and
certain types of viruses.

Interesting effects are also found in the study of influence of
chiral nanoparticles \cite{ref6} and nanoparticles with negative
refraction index \cite{ref7} on electric dipole radiations of atoms
and molecules. Full quantum-electrodynamics theory of influence of
nanospheres of various composition including media with negative
refraction index on radiation of electric and magnetic dipoles was
developed in \cite{ref8}. However, the interference effects between
these dipoles were not investigated there.

\begin{figure}[there]
\begin{center}
\includegraphics[width=6.0cm]{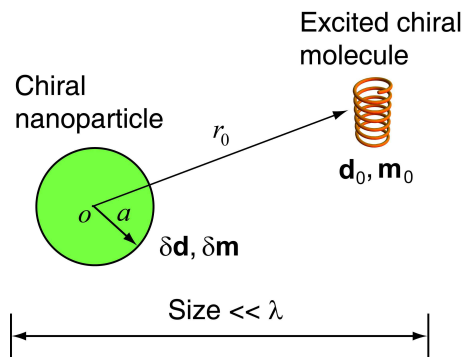}
\end{center}
\caption{\label{fig1} \textbf{Geometry of the considered problem.}
An optically active (chiral) excited molecule is placed near chiral
spherical nanoparticle having radius $a$. The total size of the
region of molecule and nanobody is much less than the wavelength
$\lambda$. Molecule is placed at the distance $r_0$ from the center
of co-ordinates, which coincides with the nanoparticle center, and
characterized by its electric (${\mathbf {d}}_{0}$) and magnetic
(${\mathbf {m}}_{0}$) dipole moments which induce 
corresponding dipole moments $\delta {\mathbf {d}}$ and $\delta
{\mathbf {m}}$ of nanoparticle.}
\end{figure}

Even more interesting problem is to investigate the possibility of
controlling radiation of optically active molecules \cite{ref9}. As
far as we know this problem has not been investigated until now. The
goal of present work is to investigate the influence of chiral
nanoparticles on spontaneous emission of optically active molecules
and to show that it is possible to discriminate drug enantiomers
with pure optical methods. Geometry of the problem is shown in
Fig.~\ref{fig1}.

As the model of an optically active molecule we assume that its
radiation is of both  electric dipole and  magnetic dipole
nature, i.e. that the Hamiltonian of molecule interaction with field
has the form \cite{ref10}

\begin{equation}
\hat {H}_{int} = - {\mathbf {\hat {d}}}{\mathbf {E}} - {\mathbf
{\hat {m}}}{\mathbf {H}}. \label{eq1}
\end{equation}

Matrix elements of operators will be written as ${\mathbf {d}}_{0}$
and $- i{\mathbf {m}}_{0} = - i\xi {\mathbf {d}}_{0}$ due to the
fact that the magnetic moment operator is purely imaginary.
Parameter $\xi$ characterize relative value of magnetic dipole
moment. Within this model, we will refer to a molecule for which
${\mathbf {d}}_{0}$ and ${\mathbf {m}}_{0}$ are parallel as a
``right'' molecule, whereas the molecule in which the magnetic
moment is antiparallel to the electric dipole moment will be
referred to as a ``left'' molecule. Thus, we are considering a
molecule in which the electrons are constrained to move along a
helical path. This picture also explains the choice of the phase
difference between electric and magnetic dipole moments.

To control the radiation of molecule with a particular chirality it
is necessary to create such nanoenvironment that it forces
interference between electric and magnetic dipole radiation of
molecules. Such fields can be created with the help of chiral
nanoparticles, i.e. nanoparticles, which can be described with
Drude-Born-Fedorov constitutive relations \cite{ref11}:

\begin{eqnarray}
 {\mathbf {D}} &=& \varepsilon \left( {{\mathbf {E}} + \beta \nabla \times
{\mathbf {E}}} \right); \nonumber \\
 {\mathbf {B}} &=& \mu \left( {{\mathbf {H}} + \beta \nabla \times {\mathbf {{\rm H}}}} \right)
 \label{eq2}
\end{eqnarray}

\noindent where ${\mathbf {D}}$, ${\mathbf {E}}$ and ${\mathbf
{B}}$, ${\mathbf {H}}$ are the inductions and the strengths of
electric and magnetic fields, respectively, $\varepsilon$, $\mu$
are the dielectric permittivity and the magnetic permeability of
chiral media, $\beta$ is the dimensional chirality parameter. The
dimensionless chirality parameter $\chi$ is defined by relation
$\chi = \beta k_{0}$, where $k_{0} = \omega / c$ is the wavenumber
in vacuum.

\begin{figure}[there]
\begin{center}
\includegraphics[width=6.7cm]{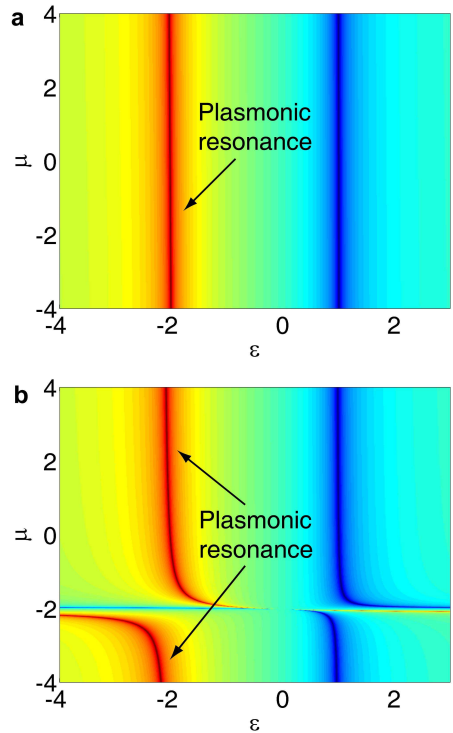}
\end{center}
\caption{\label{fig2} \textbf{Structure of TM resonances of
spherical nanoparticle.} Dependence of one of the Mie's coefficients
($n = 1$) for spherical particle \cite{ref16} as the function of
dielectric permittivity ($\varepsilon$) and magnetic permeability
($\mu$) is shown. Permittivity and permeability of the medium in
which nanoparticle is situated are equal to unit. \textbf{a},
Nanoparticle without chirality. \textbf{b}, Nanoparticle with
nonzero chirality $\chi = \beta k_{0} = 0.1$. Size parameter of
nanosphere is $k_{0} a = 0.1$, where $k_{0}$ is the wavenumber in
vacuum.}
\end{figure}

The simplest example of chiral nanoparticles is a drop of sugar
solution. Today, thanks to the development of nanotechnology it is
possible to prepare chiral nanoparticles of various forms. As an
example, one should note the artificially prepared chiral
nanoparticles in the form of segments of nanospirals \cite{ref12} or
from asymmetric clusters of spherical nanoparticles \cite{ref13}. An
important way to obtain chiral nanoparticles is the introduction of
metal nanoparticles into spiral organic molecules \cite{ref14}. In
the present work for simplicity we will consider a homogeneous
spherical nanoparticle, the substance of which is described by
constitutive equations (\ref{eq2}). The problem of scattering of
plane waves of different polarizations on chiral spherical particle
can be solved exactly by analogy with the Mie solution
\cite{ref15,ref16}. In Fig.~\ref{fig2} the structure of TM plasmonic
resonances for spherical nanoparticle of radius $a$ without
chirality and with chirality factor $\chi = \beta k_{0} = 0.1$,
constructed within the framework of solutions \cite{ref15,ref16} is
shown.

From this figure it is seen, that TM modes in nonchiral nanoparticle
have electric dipole plasmon resonance at $\varepsilon \approx - 2$,
nearly independent of magnetic permeability. Any chirality - even
small - of the nanoparticle (see Fig.~\ref{fig2}b) leads to a
radical change in the shape of the resonance line, which becomes
dependent in a non-trivial way on both the dielectric and the
magnetic permeability of the material from which the nanoparticle is
made. Below we will see that these ``hybrid'' chiral-plasmon modes
in nanoparticle provide the effective interference between radiation
of electric and magnetic dipole moments of chiral molecule.

To estimate the interaction of the radiating optically active
molecule with chiral nanoparticle we can neglect retardation effects
(see Fig.~\ref{fig1}). In this case, in the field of radiating
chiral molecules the near-field is dominating

\begin{eqnarray}
 {\mathbf {E}}_{0} &=& {\frac{{3{\mathbf {n}}\left( {{\mathbf {n}}{\mathbf {d}}_{0}}  \right) - {\mathbf {d}}_{0}}} {{r^{3}}}}; \nonumber \\
 {\mathbf {H}}_{0} &=& - i{\frac{{3{\mathbf {n}}\left( {{\mathbf {n}}{\mathbf {m}}_{0}}  \right) - {\mathbf {m}}_{0}}} {{r^{3}}}}
 \label{eq3}
\end{eqnarray}

\noindent where $r$ is the distance between the molecule and the
observation point, ${\mathbf{n}}$ is the unit vector in the
direction from the molecule to the observation point, and the
monochromatic time dependence of fields ($e^{ - i\omega t}$) is
omitted.

Fields (\ref{eq3}) induce the electric $\delta {\mathbf {d}}$ and
the magnetic $\delta {\mathbf {m}}$ dipole moments in nanoparticle:

\begin{eqnarray}
 \delta {\mathbf {d}} &=& \alpha _{EE} {\mathbf {E}}_{0} \left( {{\mathbf
{r}}_{0}}  \right) + \alpha _{EH} {\mathbf {H}}_{0} \left( {{\mathbf
{r}}_{0}}  \right); \nonumber \\
 \delta {\mathbf {m}} &=& \alpha _{HE} {\mathbf {E}}_{0} \left( {{\mathbf
{r}}_{0}}  \right) + \alpha _{HH} {\mathbf {H}}_{0} \left( {{\mathbf
{r}}_{0}}  \right) \label{eq4}
\end{eqnarray}

\noindent where $\alpha _{EE}$, $\alpha _{EH}$, $\alpha _{HE}$ and
$\alpha _{HH}$ are the tensors of electromagnetic polarizability of
spherical nanoparticle in uniform external fields and ${\mathbf
{r}}_{0}$ is the radius-vector of the molecule's position. The
possibility of applying the formula (\ref{eq4}) to strongly
nonuniform fields (\ref{eq3}) is associated with an unique property
\cite{ref17} of ellipsoids and spheres, according to which the
tensor of polarizability of an ellipsoid in nonuniform field of
dipole source is defined by usual polarizabilities and by dipole
field, which is averaged over the volume of particle. In the case of
spheres such an averaging gives value of the dipole field in the
center of sphere, i.e. ${\mathbf {E}}_{0} \left( {{\mathbf {r}}_{0}}
\right)$ and ${\mathbf {H}}_{0} \left( {{\mathbf {r}}_{0}} \right)$,
respectively.

For a spherical particle with constitutive relations (\ref{eq2}) it is easily to
find that the quasistatic electromagnetic polarizabilities are

\begin{eqnarray}
 \alpha _{EE} &=& a^{3}{\frac{{\left( {\varepsilon - 1} \right)\left( {\mu +
2} \right) + 2\varepsilon \mu \chi ^{2}}}{{\left( {\varepsilon + 2}
\right)\left( {\mu + 2} \right) - 4\varepsilon \mu \chi ^{2}}}}; \nonumber \\
 \alpha _{HH} &=& a^{3}{\frac{{\left( {\mu - 1} \right)\left( {\varepsilon +
2} \right) + 2\varepsilon \mu \chi ^{2}}}{{\left( {\varepsilon + 2}
\right)\left( {\mu + 2} \right) - 4\varepsilon \mu \chi ^{2}}}}; \nonumber \\
 \alpha _{EH} = - \alpha _{HE} &=& a^{3}{\frac{{3\chi \varepsilon \mu
i}}{{\left( {\varepsilon + 2} \right)\left( {\mu + 2} \right) -
4\varepsilon \mu \chi ^{2}}}} \label{eq5}
\end{eqnarray}

\noindent where $a$ is nanosphere radius, $\chi = \beta k_{0}$,
$\varepsilon $ and $\mu$ are dimensionless chirality parameter,
dielectric permittivity and magnetic permeability of the
nanoparticle, respectively, and it is assumed that the particle is
located in vacuum.

Since the total size of the molecule and nanoparticle system is much
less than the wavelength, the dipole momenta of molecule and
particle are added in phase and the decay rate of radiation from
such a composite system is described by

\begin{equation}
\Gamma _{rad} = {\frac{{4k_{0}^{3}}} {{3\hbar}} }{\left| {{\mathbf
{d}}_{0} + \delta {\mathbf {d}}} \right|}^{2} + {\frac{{4k_{0}^{3}}}
{{3\hbar }}}{\left| {{\mathbf {m}}_{0} + i\delta {\mathbf {m}}}
\right|}^{2} \label{eq6}
\end{equation}

\noindent where $\hbar$ is the Planck's constant.

Since the orientation of a molecule can be arbitrary relative to the
particle, to obtain an effective decay rate one should averaged
(\ref{eq6}) over the orientations of the molecule, or, equivalently,
over the unit vector ${\mathbf{n}}$. As a result we obtain

\begin{eqnarray}
\Gamma _{rad,eff} &=&{\frac{{4k_{0}^{3}{\left \vert {{\mathbf{d}}_{0}} \right\vert }^{2}}}{{3\hbar }}}\left( {1+{\frac{{2}}{{r_{0}^{6}}}}{ \left\vert {\alpha _{EE}-i\xi \alpha_{EH}}\right\vert }^{2}}\right. \nonumber \\
&&\left. {+{\left\vert {\xi }\right\vert
}^{2}+{\frac{{2}}{{r_{0}^{6}}}}{ \left\vert {i\alpha _{HE}+\xi
\alpha_{HH}}\right\vert }^{2}}\right) \label{eq7}
\end{eqnarray}

\noindent where $\xi$ defined by relation ${\mathbf {m}}_{0} = \xi
{\mathbf {d}}_{0}$.

Usually, the electric dipole moment of the molecule is much larger
than the magnetic dipole moment ${\left| {{\mathbf {d}}_{0}}
\right|} \gg {\left| {{\mathbf {m}}_{0}}  \right|}$ or $\xi \ll 1$.
Chirality parameter, even in the hypothetical metamaterials is also
small $\chi \ll 1$. Because of this, the first (electric) term in
(\ref{eq7}) is usually greater than the second (magnetic) one.
Therefore, the interference between electric and magnetic fields is
possible only when two non-trivial conditions take place.

1. The system must have chiral-plasmon resonance, that is, the
following condition should be satisfied

\begin{equation}
\left( {\varepsilon + 2} \right)\left( {\mu + 2} \right) -
4\varepsilon \mu \chi ^{2} = 0 \label{eq8}
\end{equation}

\noindent (cf. Fig.~\ref{fig2}a). Under this condition, the
contribution of magnetic radiation increases.

2. The electric dipole moment induced in the nanoparticle must be
zero, that is the following condition should take place

\begin{equation}
\alpha _{EE} - i\xi \alpha _{EH} = 0. \label{eq9}
\end{equation}

The simultaneous solution of (\ref{eq8}) and (\ref{eq9}) determines
the values of dielectric permittivity and magnetic permeability of
nanoparticle where the radiative decay rate of chiral molecules is
minimal and, therefore, the interference between electric dipole and
magnetic dipole radiations is maximal

\begin{eqnarray}
 \mu \ast &\to& - {\frac{{2d_{0}}} {{d_{0} + 2m_{0} \chi}} }; \nonumber \\
 \varepsilon \ast &\to& - {\frac{{2m_{0}}} {{m_{0} + 2d_{0} \chi}} }.
 \label{eq10}
\end{eqnarray}

\noindent Equations (\ref{eq10}) are the key equations for chiral
molecule discrimination and enantiomer selectivity.

It is very important to note that when the sign of $m_{0}$ is
changed, i.e. when changing the chirality of the molecule the
``resonance'' magnetic permeability varies slightly and remains
approximately equal to $-2$. On the other hand, the "resonance"
permittivity has different signs for the molecules of different
chirality. This means that both nanoparticles with simultaneously
negative $\varepsilon$ and $\mu$ (DNG metamaterials), and
nanoparticles with negative $\mu$ and positive $\varepsilon$ (MNG
metamaterials or magnetic plasma) are suitable for the effective
control of the radiation of chiral molecules. From point of view of
practical implementation of such control the MNG nanoparticles seem
to be the most suitable , because they can be realized within
well-developed technology of Split Rings Resonators
\cite{ref18,ref19}.

\begin{figure}[there]
\begin{center}
\includegraphics[width=6.5cm]{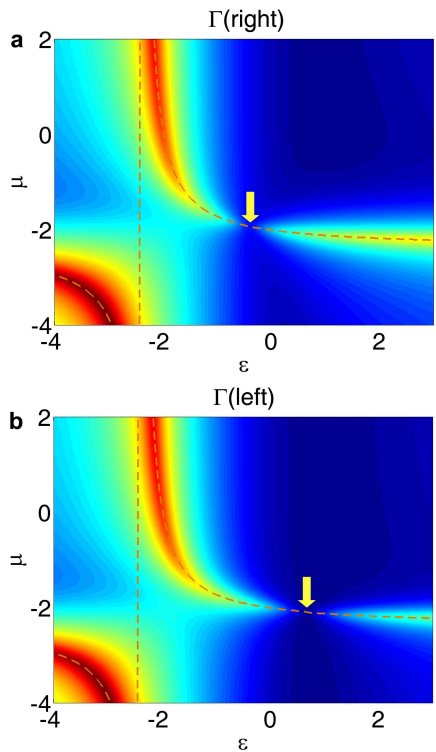}
\end{center}
\caption{\label{fig3} \textbf{Effective radiative decay rate of
chiral molecule placed near chiral nanoparticle.} Effective
radiative decay rate (see equation (\ref{eq7})) is shown as the
function of $\varepsilon $ and $\mu $. \textbf{a}, The case of
``right'' molecule with $m_{0} / d_{0} = \xi = + 0.1$. \textbf{b},
The case of ``left'' molecule with $m_{0} / d_{0} = \xi = - 0.1$.
Spherical nanoparticle is made from material with chirality $\chi =
0.2$; imaginary parts of dielectric permittivity and magnetic
permeability are ${\varepsilon} '' = {\mu} '' = 0.1$. The dashed
line corresponds to position of the chiral-plasmon resonance
(\ref{eq8}). Yellow arrow indicates the minimum rates on the
resonance curve. Molecule is placed close to nanoparticle surface.}
\end{figure}

Fig.~\ref{fig3} shows the decay rates of the ``right''
(Fig.~\ref{fig2}a) and ``left'' (Fig.~\ref{fig2}b) molecules on the
dielectric permittivity and magnetic permeability of the material
from which the nanoparticle is made.

This figure shows that, indeed, for the values of dielectric
permittivity and magnetic permeability, as determined by the
equation (\ref{eq10}), the decay rates of ``right''
(Fig.~\ref{fig3}a) or ``left'' (Fig.~\ref{fig3}b) molecules are
close to zero. It is this fact that determines the possibility of
their separation.

\begin{figure}[there]
\begin{center}
\includegraphics[width=6.7cm]{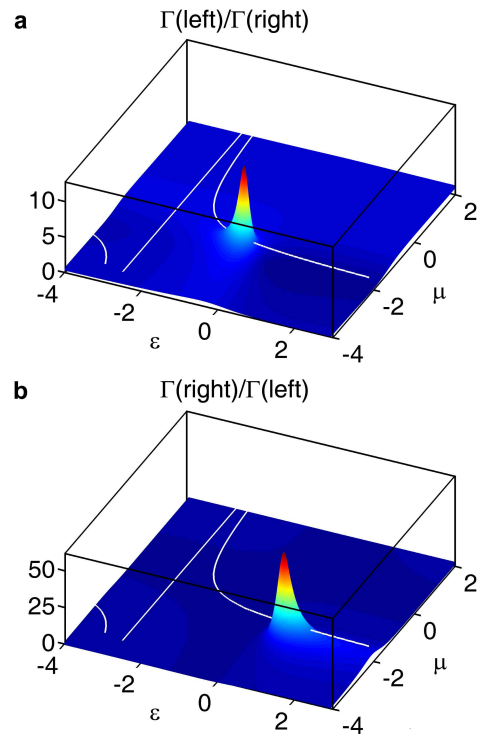}
\end{center}
\caption{\label{fig4} \textbf{Relation between effective radiative
decay rates of molecules with different chirality.} \textbf{a}, The
ratio of effective radiative decay rate (see equation (\ref{eq7}))
of the ``left'' molecule with $\xi = - 0.1$ to the radiative decay
rate of the ``right'' molecule with $\xi = + 0.1$. \textbf{b}, The
ratio of effective radiative decay rate (see equation (\ref{eq7}))
of the ``right'' molecule with $\xi = + 0.1$ to the radiative decay
rate of the "left" molecule with $\xi = - 0.1$. The dependence on
$\varepsilon$ and $\mu$ is presented. Spherical nanoparticle is made
from material with chirality $\chi = 0.2$, imaginary parts of
dielectric permittivity and magnetic permeability are ${\varepsilon
}'' = {\mu} '' = 0.1$. The white line shows the position of
chiral-plasmon resonance of nanoparticles (see equation (\ref{eq8})
and Fig.~\ref{fig3}). Molecule is placed close to nanoparticle
surface.}
\end{figure}

To determine a quantitative measure of discrimination, the ratio of
the decay rate of the ``left'' molecules to the decay rate of the
``right'' ones and vice versa is shown in Fig.~\ref{fig4}.

This figure shows that, indeed, for the chosen parameters, there is
a difference in the radiative decay rates of ``right'' and ``left''
molecules up to 10 or 50 or more times depending on the chirality of
molecule considered as a reference one. In other words, the
nanoparticles with parameters (\ref{eq10}) will enhance radiation of
the ``right'' molecules and slow down radiation of the ``left''
molecules, and vice versa. Let us stress that ``left-handedness'' of
chiral sphere (DNG material) or its negative $\mu$ (MNG material)
are of crucial importance for such discrimination.

Until now we have considered the usual case when the electric dipole
momentum of molecule is greater than the magnetic dipole momentum,
that is the case when $\xi \ll 1$. In hypothetic case when the
magnetic dipole momentum is dominating, that is, in the case where
$\xi \gg 1$ it follows from equation (\ref{eq9}) that one needs
making use of DNG or $\varepsilon$ negative (ENG) materials to
obtain enhancement of radiation of ``right'' or ``left'' molecule.

The proposed theory allows one to estimate the parameters of
nanoparticles required for the discrimination of the radiation of
``left'' or ``right'' molecules. It is very important that, despite
the simple character of the proposed theory, it is accurate enough.
The accuracy of the present approach directly follows from the
comparison with the results of full quantum-electrodynamics
calculations performed by us without any approximations
\cite{ref20}.

\begin{figure}[there]
\begin{center}
\includegraphics[width=6.5cm]{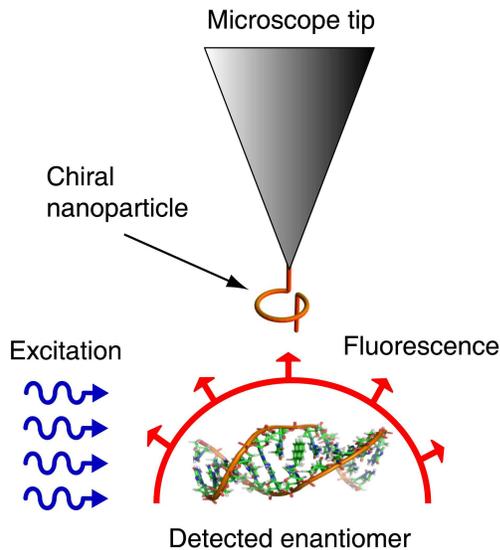}
\end{center}
\caption{\label{fig5} \textbf{Operation principle of the nanoscope
to distinguish ``right'' and ``left'' enantiomers.} Chiral
nanoparticle (or short nanospiral) is placed on the tip of the
microscope. Such nanoparticle is specially tuned to satisfy
conditions (\ref{eq10}). In such a way miscroscope is tuned to
accelerate the radiation of a molecule with selected (``left'' or
``right'') chirality depending on the chirality of the nanoparticle.
Excitated molecule with unknown chirality is radiating and its
radiation is detected (or not detected) by the microscope, depending
on the chirality of molecule.}
\end{figure}

The predicted effect of influence of chiral nanoparticles on the
radiation of optically active molecules can be used in many
applications. Here we mention only two, the most obvious ones. The
first application is observing separately ``right'' or ``left''
molecules with a scanning microscope (see Fig.~\ref{fig5}).

To do so, specially tuned (see equation (\ref{eq10})) nanoparticles
are attached to the tip of a scanning microscope to increase the
spontaneous emission of a molecule with a selected chirality. A
short metallic spiral can be used as such nanoparticle, because it
can have required chirality and negative magnetic response. Indeed,
the intensity of the fluorescence of a molecule is defined by
relation

\begin{equation}
I_{fluor} = \hbar \omega {\frac{{\Gamma _{pump} \Gamma _{rad}}}
{{\Gamma _{pump} + \Gamma _{nonrad} + \Gamma _{rad}}} } \label{eq11}
\end{equation}

\noindent where $\Gamma _{pump}$, $\Gamma_{rad}$ and $\Gamma
_{nonrad}$ are the rate of excitation, radiative and nonradiative
decays of a molecule, respectively. In the presence of nanoparticles
with the parameters (\ref{eq10}) the radiative decay rate of one
sort of molecules (e.g. ``right'' ones) increases, while the
radiative decay rate of molecules with opposite chirality (``left'')
is inhibited. Due to this fact, the contrast between the brightness
of the ``right'' and the ``left'' molecules can reach value of
10-100 or more times and, consequently, one will see images of
molecules with chosen chirality only.

\begin{figure*}[there]
\begin{center}
\includegraphics[width=12.0cm]{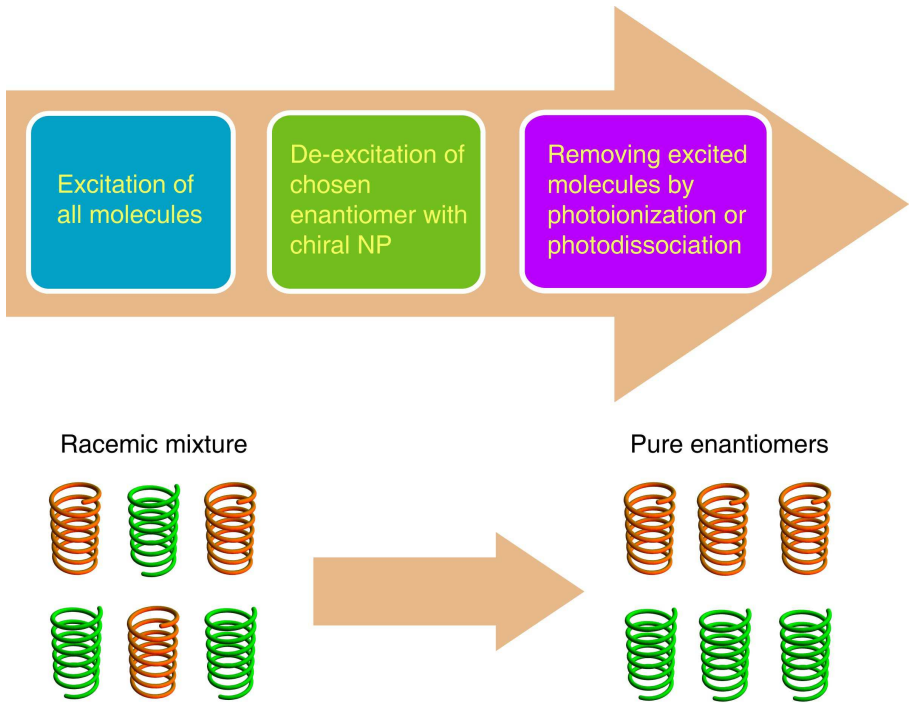}
\end{center}
\caption{\label{fig6} \textbf{Operation principle for separating of
``right'' and ``left'' enantiomers.} The racemic mixture with
excited enantiomers of both types of chirality is placed in the
specially created reaction chamber. Effective interaction of the
chiral nanoparticle which is specially tuned (according to
conditions (\ref{eq10})) takes place only with molecules with
selected chirality. It allows them quickly became unexcited, whereas
molecules with another type of chirality remaining excited. Such
molecules can be removed from the chamber by different methods. As a
result, the desired pure enantiomer will be accumulated in the
chamber.}
\end{figure*}

Significantly more important application of our results is the
separation of enantiomers in racemic mixtures. The operation
principle of such a device is shown in Fig.~\ref{fig6}.

The key element in this scheme is the reaction chamber, which
contains nanoparticles prepared in accordance with equation
(\ref{eq10}). The racemic mixture of enantiomers is placed in this
chamber and then excited by one or another way (e.g.
photoexcitation). Due to presence of chiral nanoparticles one type
of the optically active enantiomers radiates efficiently and goes to
ground state quickly, while the remaining excited enantiomers can be
ionized by a resonant field, and then removed from the chamber.
Other methods of removal of excited molecules or their decay
products are also possible. As a result, the desired pure enantiomer
will be accumulated in the chamber. Our approach for pure optical
spatial separation of enantiomers has some advantages in comparison
with other proposal \cite{ref21}, because we have no need to use
very low temperatures.

In conclusion, the theory of radiation of optically active molecule placed
near chiral nanoparticles is elaborated. It is shown that for special
parametres of nanoparticle (DNG, MNG or ENG metamaterials) it is possible to
increase substantially radiation of ``right'' molecules and suppress
radiation of ``left'' molecules and vice versa. An application of this
effect to separate racemic mixtures of drug enantiomers is proposed.

\newpage \textbf{Acknowledgements} \\ We are grateful to D. Bloch for
fruitful discussions of this topic. This work has been supported by
the Russian Foundation for Basic Research (grants 11-02-91065,
11-02-92002, 11-02-01272) and performed in the frame of
International program of Scientific Cooperation (``P.I.C.S.'' No 5813)
between C.N.R.S. and the Russian Foundation for Basic Research.

\bigskip
\textbf{Author contributions} \\ All authors have contributed to
this paper and agree to its contents.

\bigskip
\textbf{Competing interests} \\ The authors declare that they have
no competing financial interests.

\bigskip
\textbf{Correspondence} \\ Correspondence and requests for materials
should be addressed to V.K. (email: vklim@sci.lebedev.ru).

\end{document}